\documentstyle[preprint,aps,epsfig]{revtex}
\tightenlines
\begin{document}
\title{Dimensionality effects on the Holstein polaron}
\author{Li-Chung\ Ku$^{1,2}$, S.\ A.\ Trugman$^1$,
and J.\ Bon\v ca$^3$}
\address{
$^1$Theoretical Division, Los Alamos National Laboratory,
    Los Alamos, New Mexico 87545, U.S.A.\\
$^2$Department of Physics, University of California,
    Los Angeles, California 90024, U.S.A. \\
$^3$FMF, University of Ljubljana and J. Stefan Institute, 1000,
Ljubljana, Slovenia}
\date{\today}
\draft
 \maketitle
\begin{abstract}
Based on a recently developed variational method, we explore the properties of
the Holstein polaron on an infinite lattice in $D$ dimensions, where 
$ 1 \le D \le 4$. The computational method converges as a power law, 
so that highly accurate
results can be achieved with modest resources. We present the
most accurate ground state energy (with no small parameter) to date for
polaron problems, 21 digits for the one-dimensional (1D) polaron at
intermediate coupling. The dimensionality effects on polaron band dispersion,
effective mass, and electron-phonon (el-ph) correlation functions are
investigated in all coupling regimes. It is found that the crossover to large
effective mass of the higher-dimensional polaron is much sharper than the 1D
polaron. The correlation length between the electron and phonons decreases
significantly as the dimension increases. Our results compare favorably with
those of quantum Monte Carlo, dynamical mean-field theory, density-matrix
renormalization group, and the Toyozawa variational method.  We demonstrate that
the Toyozawa wavefunction is qualitatively correct for the ground state energy
and the 2-point electron-phonon correlation functions, but fails for the
3-point functions.  Based on this finding, we propose an improved 
Toyozawa variational wavefunction.
\pacs{PACS numbers: 74.20.Mn, 71.38.+i, 74.25.Kc}
\end{abstract}

\section{Introduction}

The Holstein model, as a paradigm for polaron formation, has attracted renewed
interest in recent years because several lines of experimental evidence support
the presence of polaron carriers in strongly correlated electronic materials, 
including colossal magnetoresistance manganites \cite{zhao}, 
organics \cite{campbell}, 
quasi-1d systems, and high-$T_c$ cuprates \cite{Htc,zxshen}. 
Theoretical research on polaron
physics began six decades ago, and the problem remains unsolved due to its
intrinsic many-body complexity from the electron-phonon interaction. 
(The problem of excitons coupled to phonons is formally equivalent 
\cite{rashba}.)  Standard
perturbation treatments \cite{Holstein,appel} are usually limited to a
particular parameter regime. With constantly growing computational resources,
various numerical techniques have been applied to polaron problems in recent
years, which give the most reliable results in the physically interesting
crossover regime. These techniques include finite-cluster exact diagonalization
(ED)\cite{alex94,wellein96,mello97,fehske97,capone97,white99}, quantum Monte
Carlo (QMC)\cite{korn98,korn99}, density-matrix renormalization group
(DMRG)\cite{DMRG}, and the global-local variational method (GLVM)\cite{GL}.

Recent numerical studies have focused on the 1D lattice model. Due to the
enormous phonon Hilbert space in three dimensions, the dimensionality effects
on the polaron problems are less studied except in the adiabatic (or
semiclassical) approximations\cite{adiab1,adiab2}, and in perturbation
theory\cite{romero00}. QMC is also capable of computing the energy and
effective mass of the 3D polaron, 
but the full dispersion $E( \vec k )$ is only
reliable in the strong-coupling
regime. However, with a recently developed variational method, we can compute
the polaron effective mass, band dispersion, and el-ph correlation functions of the
ground and low-lying excited states in all coupling regimes, preserving the full
quantum dynamical feature of phonons (details can be found in Ref.
\cite{bonca99}). The variational space is defined on an infinite lattice, although
only a finite separation is allowed between the electron
and the surrounding phonons in current implementations. 
We systematically expand the variational space so
that highly accurate results can be achieved with modest computational
resources.

The main purpose of this paper is to characterize the Holstein polaron in
higher dimensions. We consider a single-electron Holstein Hamiltonian on a
D-dimensional hypercubic lattice,
\begin{eqnarray}
H &=& H_e + H_{\mbox{\scriptsize {\it el-ph}}} + H_{ph} \nonumber \\
  &=& - t\sum\limits_{<i,j>} {(c_i^\dag c_{j}  + h.c.) -
\lambda \sum\limits_j {c_j^\dag c_j (a_j  + a_j^\dag ) + \omega \sum\limits_j
{a_j^\dag a_j } } }
~,  \label{eq:ham}
\end{eqnarray}
where $c_j^\dag $ creates an electron and $a_j^\dag $ creates a phonon on site
$j$. The parameters of the model are the nearest-neighbor hopping integral $t$,
the el-ph coupling strength $\lambda$, and the phonon frequency $\omega$.  The
electron is coupled locally to a dispersionless optical phonon mode
\cite{com1}. There are two commonly defined dimensionless control parameters,
the adiabaticity ratio $\gamma = \omega/t$, and the el-ph coupling
strength $\alpha = E_p / 2Dt$, which is defined as the ratio of polaron energy
for an electron confined to a single site $E_p = \lambda^2/\omega $, and 
the free electron half bandwidth $2Dt$.  The strong (weak) coupling regime refers
to $\alpha > 1 ~(< 1)$, and the adiabatic (antiadiabatic) regime refers to $ \gamma
< 1 ~ (> 1) $. 
An additional dimensionless parameter is $g = \lambda / \omega$,
which appears in strong-coupling perturbation theory.

A variational space is constructed beginning with a root state,
the electron at the origin with no phonon excitations,
and acting repeatedly with the off-diagonal terms 
($t$ and $\lambda$) in the Hamiltonian,
Eq.\ \ref{eq:ham}.  States in generation $m$ are those
that can be created by acting $m$ times with off-diagonal terms.
All translations of these states on an infinite
lattice are included, and the problem is diagonalized 
for a given momentum $\vec k$ using a Lanczos method \cite{bonca99}.
A small variational space, with 7 states per electron site (unit cell),
is shown in Fig. \ref{fig:tight}.  (The more accurate numerical computations
are done with over $10^7$ states per unit cell.)

The total number of states $N_{st}$ per unit cell after $m$ generations increases
exponentially, as $(D+1) ^ m$, where $D$ is the spatial dimension.
(The bipolaron has the same $(D+1) ^ m$ dependence, but
with a larger prefactor.)  The perhaps surprising fact is
that while the size of the Hilbert space grows exponentially with $m$,
the error in the ground state energy decreases exponentially,
because states are added in a fairly efficient order.
Figure \ref{fig:convg} shows the fractional error in the ground
state energy as a function of the number of basis states in Hilbert space.
The accuracy is determined by comparing the energy 
as the size of the Hilbert space is increased.
At intermediate coupling in
any dimension, the energy improves by about a factor of 8
with each generation  \cite{strongc}.
In 1D, each added generation approximately doubles the size of 
the Hilbert space,
whereas in 4D, the size increases five-fold.
This rapid convergence at intermediate-coupling is valuable since no analytic
approach is reliable in this regime.  
Table I lists the
energies for 1D to 4D polarons at intermediate- to weak-coupling. The accuracy,
21 digits for 1D polaron, is high compared to that of other numerical
methods, such as 2 or 3 digits for QMC, 6 digits for DMRG (or GLVM), and up to
8 digits for ED \cite{fehskep}. Moreover, for the 3D polaron at intermediate-
to strong-coupling, an energy accuracy of 8-10 decimal places can be achieved
in the nonadiabatic regime with fewer than $3 \times 10^6$ basis states.
To obtain an accuracy beyond 13 digits, the code is executed in
quadruple precision. The present variational method requires only power-law
time to achieve a given accuracy (in any dimension), 
which is a qualitative improvement on
exact diagonalization as it is currently implemented, the latter requiring
exponential time.

In this paper we present detailed studies of the dimensionality effect on the
Holstein polaron. First of all, we explore the polaron characteristics in the
$k=0$ ground state, and compare our results with previous studies from QMC,
DMRG, and dynamical mean-field theory (DMFT). Secondly, we compute the el-ph
correlation function and the polaron energy $E(\vec k)$.  Finally, the validity of the
Toyozawa variational method is investigated by calculating the ground-state
energy, and the 2-point and 3-point el-ph correlation functions.


\section{Small-polaron crossover}
\subsection{Quasiparticle weight $Z_k$ and effective mass $m^*$}
The small polaron crossover or ``self-trapping transition'' has been one of the
core issues in polaron problems. Adiabatic theory suggests that the
polaron in 2D and 3D (but not in 1D), is in an ``extended'' state with
an infinite radius below an el-ph coupling threshold $\lambda_c$, and beyond which
is a ``localized'' state with infinite effective mass. (This phenomenon is
usually termed the ``self-trapping transition''.)  However, our studies confirm
that in all dimensions, there is a crossover rather than a self-trapping transition 
(ground state properties are analytic), if the parameters are
finite. This result is consistent with some other recent studies \cite{romero00,korn98}
and corroborates the theorem of Gerlach and L\"owen \cite{gerlach}. 

The quasiparticle weight (renormalization factor) is
defined by the overlap (squared) between the bare electron and a polaron, i.e.,
\begin{eqnarray}
Z_k= \left| \langle \Psi_{0,k}|c^\dag_k| 0 \rangle \right|^2 ,
\end{eqnarray} 
in which $|\Psi_{0,k}\rangle$ is the ground state wavefunction
of a polaron and $| 0 \rangle$ is the vacuum state. 
$Z_k$ can be measured in photoemission or tunneling experiments.
Figure \ref{fig:crossover}
shows the crossover of $Z_{\vec k=0}$
as a function of $1/\alpha$ at $g=3$, for 1D to 4D
cases. We see that {\em the crossover to large effective mass of the
higher-dimensional polaron is much sharper than the $1D$ polaron}. For $D>1$,
a fairly abrupt crossover occurs at $\alpha > 1$, whereas the crossover for the 1D
polaron spans a wide range of $\alpha$.  With a smaller $g$ (but greater than
$1$), the crossover will be slower but with the same dimensional
characteristics.  In the limit $1/\alpha \to 0$, the 
phonon wavefunction contracts to the electron site, with
$Z_k = \exp (-g^2)$. 
The inset shows a comparison of $Z_k$ and $m_0/m^*$ for the 1D
polaron. Their fractional difference $\delta$, defined as $(m_0/m^* -Z_k)/Z_k$, 
is shown
as a dotted line. The maximum $\delta$ is $22\%$, 
in the intermediate
coupling regime, while the minimum occurs as
$1/\alpha \to 0$ (small $t$), where $\delta$ is the order of $t^2$ from
strong-coupling perturbation theory (SCPT). We find that $\delta$ decreases
significantly as the dimension increases. The maximum
difference $\delta_{max}$ is $4.5\%$ for the 2D, and $2.0\%$ for the 3D
polaron. For $g=\sqrt{5}$, $\delta_{max}$ in 3D
drops to $0.63\%$.

The ground state energy $E$
satisfies $ E = \epsilon _0 -2 t \cos (k) + \Sigma(k,E)$,
where $\Sigma(k,E)$ is the self-energy.
$Z_{k=0}$ is the probability
of the wavefunction on the root site, and from first
order perturbation theory $Z_{k=0} = \partial E / \partial \epsilon _0$,
resulting in $ Z_{k=0} = 1/[1- \partial \Sigma(k,E) / \partial E ] $.
The origin of the difference between the inverse mass and $Z_{k=0}$
lies in the $k$-dependence of the self-energy,
\begin{eqnarray}
{m_0 \over {m^*}} - Z_{k=0} ~ = ~
{1 \over {2t}} {{\partial ^2 \Sigma(k,E)} \over {\partial k^2}} ~ \big/ ~
\big( 1-   { {\partial \Sigma(k,E)} \over {\partial E} } \big) ~ ,
\end{eqnarray}
where the derivatives are evaluated at the ground state
energy $E = E_0$ and $k=0$.
In the variational space of Fig.\ \ref{fig:tight} or in the full space,
the self-energy has nonzero $k$-dependence because distinct
unit cells are connected at branch level (path $1-2-3 \dots$),
in addition to the trivial connection at root level.
A restricted variational space, the comb basis,
allows phonon excitations only on
the electron site, as shown in Fig.~\ref{fig:comb}.
In this subspace, the self-energy is k-independent,
since the only path between unit cells is at the root level.
The self-energy remains k-independent even in a larger
space in which the tree trunks sprout lateral branches,
so long as the branches do not connect to neighboring unit cells.
For these cases, the $Z$ factor and inverse mass are
identical, $\delta =0$.
In ${\cal O}(t)$ SCPT, 
$\delta$ vanishes for the same reason and $Z_{k=0} =m_0/m^*=\exp (-g^2)$.


The effect of dimensionality on $\delta $ is made plausible by the
following. In the Holstein model, the dimensionality D does not directly
affect the term $H_{\mbox{\scriptsize {\it el-ph}}}$ in Eq.\ 1, because the
el-ph coupling is local and the phonon is dispersionless. 
High dimensional polarons share the same simplicity of the el-ph coupling as
1D.  Furthermore, we see (in next section) that the
el-ph correlation length decreases as D increases. 
Thus the k-dependence of the self-energy weakens in higher dimensions.
The above arguments do not, however, hold
for the Fr\"ohlich model (or the extended Holstein
model) with longer-range el-ph coupling \cite{india,EHM1,EHM2}, where $Z_k$
and $m_0/m^*$ behave quite differently.

\subsection{Comparison with QMC, DMRG and DMFT}
Figure \ref{fig:DMRG} shows our results for the effective mass as a function of
$\alpha$ (at fixed $\gamma$ = 1.0) in comparison with DMRG and QMC. Our
results are accurate to at least four digits, which is
well below the linewidth. In all cases, Fig.\ \ref{fig:DMRG}(a)-(c), 
$m^*/m_0$ increases slowly when $\alpha$ is small, followed by a
rapid increase when $\alpha > 1$. Since it is calculated at $\omega=t=1.0$ (not
a small t), the mass behaves differently than $\exp( g^2 )$ that SCPT suggests.
Note that the crossover is more rapid as D increases, which is consistent
with the results in previous section.  In every dimension, our results are in
quantitative agreement with QMC. The numerical error in QMC is approximately
$0.1\%$ to $0.3\%$\cite{korn98}, which is good though less accurate
than finite cluster ED or the present approach.
DMRG is
generally considered a powerful tool in dealing with many-body problems.
Using DMRG, Jeckelmann and White have calculated Holstein polaron properties
in 1D and 2D.  DMRG seems to be most successful calculating the 
ground state energy (at $k=0$) and el-ph correlation functions.
However, finite-size scaling is required for DMRG to compute 
$m^*$ \cite{DMRG}, which becomes more difficult for $D > 1$. In 1D, Fig.\
\ref{fig:DMRG}(a), the results from DMRG are as accurate as QMC.
DMRG does not, however, calculate the mass accurately in 2D,
Fig.\ \ref{fig:DMRG}(b).


Dynamical mean-field theory has previously been applied to the Holstein polaron
problem\cite{DMFT}. The approach is exact in infinite dimensions but an
interpolation to 3D lattices is made possible by using a semielliptical free
density of states $N(E)$ to mimic the low-energy features. Figure \ref{fig:DMFT}
shows a comparison of our results on a cubic lattice to DMFT, 
which is made by setting the bandwidths equal.
Overall, in panel (a), we see a qualitative agreement
between the two calculations.  DMFT is accurate in the
strong-coupling regime, where the surrounding phonons are predominately on
the electron site. This is also the regime where strong-coupling
perturbation theory works well. 
In Fig.\ \ref{fig:DMFT}(b), we
see our numerical results in agreement with weak-coupling perturbation
theory in $\lambda$.  
However, DMFT fails to compute $m^*$ correctly in the weak-coupling
regime. The reason is that in DMFT, the lattice problem is mapped onto a
self-consistent local impurity model \cite{DMFT2,DMFT3}, which preserves the
interplay of the electron and the phonons only {\em at the local level}. We
will see that the spatial extent of the el-ph correlations increases as the el-ph
coupling decreases, which explains the significant discrepancy in the
weak-coupling regime. It is also worth noting that DMFT neglects the $k$
dependence of self-energy, i.e., the inverse effective mass is always equal to the
quasiparticle weight. As we have pointed out above, the difference between
$m_0/m^*$ and $Z_k$ is not negligible in the intermediate- to weak-coupling
regime.

\section{electron-phonon correlations}
Next, we compute the correlation function between the electron and the phonon
displacement (lattice deformation) in the ground state,
\begin{eqnarray}
\chi(i-j) = \langle \Psi_0 | c^\dag_i c_i (a_j + a^\dag_j )|\Psi_0 \rangle.
\end{eqnarray}
This correlation function can be considered as a measure of the polaron
size\cite{romero99}. It should not be confused with the ``polaron radius'' in the
extreme adiabatic limit, which refers to the spatial extent of a
symmetry-breaking localized state. We would like to emphasize that a comprehensive
study of el-ph correlation for the ground state of the 3D polaron has not yet
been reported by any other modern numerical technique\cite{deRaedt}. The
{\em on-site correlation} has been studied by DMFT and the results are 
compared in Fig.~\ref{chi0} ~\cite{norm}. The on-site lattice distortion
$\chi(0)$ is shown as function of $\alpha$ and the rest of parameters are the same
as Fig.~\ref{fig:DMFT}. In Fig.~\ref{chi0}, $\chi(0)$ is normalized to 1 when 
$\alpha$ is infinite (i.e. $t\to 0$) according to 
$\mathop {\lim}\limits_{t\to 0} \chi(0) = 2g$.  Again, we obtain
good qualitative agreement.  The curves show an abrupt change
in slope only for large $g$, where the discrepancy with DMFT is largest.

Figure \ref{fig:deform} shows the effect of
dimensionality on the correlation function $\chi(i-j)$. 
In the strong-coupling
regime, panel (a) shows, in every dimension, a sharp drop on the first two
sites and an exponentially decaying tail. For the 3D polaron at a
distance of 3 lattice sites, $\chi(3)/\chi(0)$ drops below $10^{-4}$.
In the weak-coupling regime, panel (b),
$\chi$ has nearly a simple exponential decay with a less steep slope, which implies a
nontrivial extent of the el-ph interplay in space. In both panels, we
observe a common trend that $\chi$ decays more rapidly as the lattice dimension
increases, i.e., the surrounding phonons are more localized near the electron
in higher dimensions.  This feature enables DMFT to give
sensible results in higher finite dimensions. We have also investigated 
other 2-point el-ph
correlation functions such as $\langle c_i^\dag
c_i a_j^\dag a_j \rangle$ (not shown), 
which has dimensional characteristics similar to $\chi$.

The rapid decay of the el-ph correlation function for the higher-dimensional
polaron suggests that the off-site el-ph interplay is relatively weak in large
D. One would then expect the comb basis of Fig.\ \ref{fig:comb}, 
a subspace of the full
Hilbert space, to give a better approximation in large D. 
We check this assumption by
numerically calculating the fraction of the probability density in the 
exact ground state
that resides in the comb subspace,
\begin{eqnarray}
P_{comb} = \langle \Psi _0 | \hat P | \Psi _0 \rangle ,
\end{eqnarray}
where $\hat P$ is the projection operator onto the comb subspace and
the wavefunction $\Psi_0$ is obtained in the full variational space.
Figure \ref{fig:P_comb} shows $P_{comb}$ as a function of the inverse bare
coupling constant $1/\alpha$ for 1-4D cases. In both of the limits 
$\alpha=0$ or $\alpha=\infty$, $P_{comb}$ goes to 1. The minimum overlap
occurs in the crossover regime. As expected, $P_{comb}$ gets 
closer to 1 as D increases. For the 3D polaron, the minimum of $P_{comb}$ is
91.1\%, in contrast to 45.8\% for 1D. 
These trends can also be seen analytically.
In the adiabatic limit ($\omega=0$), perturbing in $t$
from a self-trapped state with energy $E_p$,
the self-trapping transition occurs at $\alpha=1-{1\over 4D}$.
The leading order correction of
$P_{comb}$ for the self-trapped polaron state is
\begin{eqnarray}
\Delta_{comb} &\equiv& 1-P_{comb} \nonumber \\
              &=& { 1 \over {8D\alpha ^2}} .
\end{eqnarray}
In the non-adiabatic limit, $\Delta_{comb}$ can be calculated
by SCPT to second order in the hopping $t$. It takes the following form:
\begin{eqnarray}
\Delta_{comb} &=& {g^4 e^{-2g^2} \over {2D\alpha^2 }} \sum_{n=0}^{\infty}
\sum_{m=1}^{\infty}\frac{g^{2(n+m)}}{n!\;m!} \frac{1}{(n+m)^2}.
\end{eqnarray}
The above expressions show that for a given $\alpha$ and $g$, the discrepancy
$\Delta_{comb}$ decreases as D increases and eventually vanishes in infinite D.
The comb basis should thus give a good
account for the Holstein problem in large D.
We see in Fig. \ref{fig:crossover}, however, that dimension 3 is not high
enough for the comb to give quantitatively accurate results,
and that dimension 4 is not much better.  Convergence
to higher dimensions is slow.

\section{Energy Dispersion $E(\vec k)$}
Most of the recently developed numerical methods are capable of computing the
polaron band dispersion in 1D. For the 2D polaron, the only non-perturbative
calculations of band dispersion published so far were computed by
finite-cluster ED \cite{fehske97} and Quantum Monte Carlo \cite{korn99}. Due to
the huge phonon Hilbert space in high dimensions, the previous ED results are
limited to small clusters, so that the band dispersion can only be evaluated at
a few $\vec k$ points. The QMC allows calculation of energy at any desired 
$\vec k$ point,
but is limited to the condition that the polaron bandwidth is much smaller than
the phonon frequency, which corresponds to the strong-coupling regime.

The present variational approach, however, is not subject to
any of the above restrictions \cite{bonca99}. 
Figure \ref{fig:band}(a) shows the evolution of
the band dispersion for the 3D polaron along symmetry directions in the
Brillouin zone at various el-ph coupling constants $\lambda$. Figure
\ref{fig:band}(b) shows the corresponding $Z_k$. Starting with weak coupling
$\lambda = 2.0$ (dashed line), the polaron band is close to the bare
electron band at lower band edge. The deviation between them increases as $\vec k$
increases.  When $E(k)-E(0)$ approaches $\omega$, we observe a band flattening
effect, similar to the 1D and 2D cases, accompanied by a sharp drop of
quasiparticle weight $Z_k$. 
The large $k$ lowest energy state can be
considered roughly as ``a $k=0$ polaron ground state'' plus ``an itinerant
(or weakly-bound) phonon with momentum $k$''. It is the phonon that carries the
momentum so as to make $Z_k$ essentially vanish and give a bandwidth
$E(\pi)-E(0) =\omega$. Due to the large extent of the el-ph correlations in
the flattened band, our results are less accurate in the flattened
regime \cite{accuracy}. In the case of intermediate coupling $\lambda=3.5$, the
polaron bandwidth is narrower than the phonon frequency.  The upper
part of the band has much less dispersion than the lower part but {\em with
a substantial $Z_k$ }. This indicates a distinct mechanism for the
crossover as a function of $\vec k$.  In the case of $\lambda=4.5$, the strong
el-ph interaction leads to the well-known polaron band collapse and a
significant suppression of $Z_k$ at all $k$.

\section{Toyozawa variational method}
Four decades ago, a simple and intuitive variational approach to the 1D polaron
problem was proposed by Toyozawa \cite{toyozawa}. 
This method has been successfully applied to
various fields and revisited in a number of guises\cite{toyo2,zhao97} throughout
the years. It is generally believed to provide a 
qualitatively correct description of the polaron
ground state, aside from predicting a spurious 
discontinuous change in the mass at intermediate coupling.
We show below that although the Toyozawa wavefunction
gives a good account of the ground state energy and the 2-point functions, it fails
to correctly describe the 3-point functions.

The Toyozawa wavefunction is written as a product of coherent states,
\begin{eqnarray}
|\Psi _T (k)\rangle = \sum\limits_j {e^{ikj}} ~ c_j^\dag |{0} \rangle
~  \prod\limits_m {|z_{j + m}\rangle} ~ , \label{eq:TY}
\end{eqnarray}
where $|z_i\rangle$ is a coherent state of the phonon mode on site $i$.  In the
antiadiabatic limit $\omega/t \to \infty$, this wavefunction gives the exact
solution $c_j^\dag |{0}\rangle |z_j\rangle$, where $z_j = \lambda/\omega$
and the other $z's$ are zero.  For the general case,
momentum $k=0$, the $z's$ are real and symmetric: $z_{j+m} = z_{j-m}$. To
determine the validity of the Toyozawa wavefunction, we probe the structure of
the phonon cloud in the $k = 0$ ground state by computing the following 2-point
and 3-point el-ph correlation functions,
\begin{eqnarray}
\alpha_2(j) & \equiv & \langle c_0^\dag c_0 a_j^\dag a_j \rangle\;, \label{eq:2pt} \\
\alpha_3(j,m) & \equiv & \langle c_0^\dag c_0 a_j^\dag a_j a_{m}^\dag
a_{m} \rangle  .
\end{eqnarray}
The $z's$ in Eq.\ \ref{eq:TY} are optimized
so as to give a minimum energy. It can be proved that the optimal $z's$ 
decay exponentially as a function of el-ph separation.  
Thus it always gives purely exponentially decaying
2-point functions regardless of the el-ph coupling. This, however, is not true
of the numerically exact results.  Table II and Fig.\ 
\ref{fig:Tyz} compare Toyozawa's and the numerically exact results
for intermediate coupling \cite{linden}.
We notice that the Toyozawa wavefunction gives reasonably
accurate results for the ground state energy, 2-point functions,
and $Z_{k=0}$.   In Table
II, the fractional error in energy is about
$1\%$ (with $z_{j+1}/z_{j} = 0.35568$ and $z_0=0.57033$). However, it gives wildly
inaccurate 3-points functions. 
For example, the Toyozawa $\alpha_3(1,-2)$ is a factor of 36 too large
and $\alpha_3(1,2)$ is a factor of 2 too small.
The Toyozawa $\alpha_3(5,-6)$ is too large by 6 orders of magnitude.
This failure indicates
that the electron does not organize its surrounding phonon cloud in the way
that Toyozawa suggested. Instead, by directly analyzing the exact ground state
wavefunction, we find that the electron organizes its surrounding phonons like
a traveling salesman does, namely, the polaron favors the phonon
configuration with a shorter creation path. 
(The length of the creation path is the number of off-diagonal operations,
phonon creations and electron hops, required to create a state
from the bare electron state.  Shorter paths are favored at
intermediate or weak el-ph coupling, although more on-site phonons
can be favored at large coupling.)
For example, we have,
\[
\left| {\left\langle {\left. {\Psi _0 \left| {c_0^\dag } \right.a_0^\dag \left|
{ 0} \right.} \right\rangle } \right.} \right|   >  \left| {\left\langle
{\left. {\Psi _0 \left| {c_0^\dag } \right.a_1^\dag \left| { 0} \right.}
\right\rangle } \right.} \right|    >   \left| {\left\langle {\left. {\Psi _0
\left| {c_0^\dag } \right.a_2^\dag \left| { 0} \right.} \right\rangle }
\right.} \right| > \ldots
\] in the 1-phonon subspace and
\[
\left| {\left\langle {\left. {\Psi _0 \left| {c_0^\dag } \right.a_0^\dag
a_0^\dag \left| { 0} \right.} \right\rangle } \right.} \right|   >  \left|
{\left\langle {\left. {\Psi _0 \left| {c_0^\dag } \right.a_1^\dag a_1^\dag
\left| { 0} \right.} \right\rangle } \right.} \right|    >   \left|
{\left\langle {\left. {\Psi _0 \left| {c_0^\dag } \right.a_1^\dag a_{-1}^\dag
\left| { 0} \right.} \right\rangle } \right.} \right| > \ldots
\]
in the 2-phonon subspace. The amplitude attenuates rapidly as the
phonon-creation path increases.

The numerically exact result in Fig.~\ref{fig:Tyz}(b) shows 
that it is far more favorable
to create two phonon excitations on the same side of
the electron than on opposite sides.
Therefore, we propose
to write a polaron as a sum of
two asymmetric clouds, one extending like a comet-tail primarily off to the
right and the other extending to the left,
\begin{eqnarray}
|\Psi'_T(k)\rangle = \sum_j e^{ikj} c_j^\dag |{0}\rangle \left( \ldots |z_{j-2}
\rangle|z_{j-1} \rangle|z_{j} \rangle|z_{j+1} \rangle|z_{j+2} \rangle \ldots +
\right. \nonumber \\ \left.
\ldots |z_{j+2} \rangle|z_{j+1} \rangle|z_{j} \rangle|z_{j-1} \rangle|z_{j-2}
\rangle \ldots \right) ,\label{eq:newTY}
\end{eqnarray}
where $z_{j-m} \neq z_{j+m}$, and the normalization factor has been omitted.
The optimized (minimum energy) phonon wavefunction in Eq.~\ref{eq:newTY} is
strongly asymmetric, and in fact changes sign on one side, as shown in Table
III. The main purpose of Eq.~\ref{eq:newTY} is to investigate how the simplest
asymmetric wavefunction improves the Toyozawa method. Shore and Sander
have proposed a more complicated wavefunction $|\Psi_{SS}^{IV}\rangle$
which is a sum of the symmetric term in Eq.~\ref{eq:TY} and the two asymmetric terms
in Eq.~\ref{eq:newTY} \cite{toyo3}. 
(Asymmetric wavefunctions are also considered in Ref.~\cite{toyo2}.)
The number of independent variables in $\Psi_T, \Psi'_T$, and
$\Psi_{SS}^{IV}$ is ${1\over 2}N$, $N$, and ${3\over 2}N$ respectively,
where $N$ is the number of sites that allow phonon excitations.  The
minimum energies from the above methods are compared in Fig.~\ref{fig:TY_Shore}.
It is clear that the energies are improved as we expand the variational space,
$\Psi_T \subset \Psi'_T \subset \Psi_{SS}^{IV}$. The Shore-Sander method shows
the most substantial improvement in the crossover regime. The comparison of
other polaron properties is shown in Table II and Fig.\ \ref{fig:Tyz}. Our
trial wavefunction $|\Psi'_T\rangle$ improves the energy by 30\% and
the $k=0$ Z-factor by 50\% compared to 75\% and 66\% respectively from
Shore-Sander wavefunction.  In Fig.\ \ref{fig:Tyz}(a), Eq.\ \ref{eq:newTY} gives
a more accurate 2-point function $\alpha_2(j)$ than
the original Toyozawa.  It similarly improves the
other 2-point function $\chi_j$ (not shown). Panel (b) shows two 3-point
functions, $\alpha_3(j,j+1)$ and $\alpha_3(j,-j-1)$. Due to its symmetric
phonon cloud, the Toyozawa wavefunction must give exactly the same result for
the two 3-point functions. In contrast, the exact results show that
$\alpha_3(j,j+1) \gg \alpha_3(j,-j-1)$. 
Equation \ref{eq:newTY} gives the correct behavior 
of the two 3-point functions on
nearby sites, but loses quantitative accuracy in the tails.  
Although the Shore-Sander energy is better than that of 
Eq.~\ref{eq:newTY}, the Shore-Sander 3-point functions
are actually worse.
The simplest
attempt $ |\Psi'_T \rangle$ to correct the identified shortcomings in the
Toyozawa variational wavefunction appears to be a step in the right direction,
although not as quantitatively accurate for most properties as
variational methods with more parameters \cite{GL,toyo4}.

\section{conclusion}
In summary, we have performed extensive numerical studies of the Holstein
polaron in spatial dimensions 1 through 4. 
The numerical method used adds basis
states to the Hilbert space in
an efficient order, resulting in an error that scales as a power
of the size of the Hilbert space $N_{st} ^ {- \theta}$,
where $\theta$ is a nonuniversal exponent 
$\approx 3$ at intermediate coupling in 1D, 
and $\approx 1.6$ in 3D.
This is a qualitative improvement over standard exact diagonalization,
which requires exponential effort to achieve a given accuracy.
Using modest computational resources, we obtain
by far the most accurate polaron energies and wavefunctions
available from 1D to 4D at intermediate coupling.  

Previously, a thorough investigation of the
dimensionality effect, including correlation functions,
was out of reach of numerical methods.  The main
findings of the dimensionality effects on the the Holstein polaron are
summarized as follows:
The crossover from quasi-free to large effective mass is found
to be much sharper in higher dimensions. 
As was recognized previously, there is no symmetry-breaking 
self-trapping transition for finite parameters in any dimension,
as suggested by adiabatic theory
(although there is a phase transition in the first 
excited state \cite{bonca99}).
See also Ref.~\cite{DMFT}.
Our results for $m^*$ agree with QMC,
although there is a discrepancy with DMRG in $D>1$.
The electron-phonon correlation functions
decay significantly faster in higher than lower dimensions.  This
implies a shorter el-ph correlation length in large dimensions and leads to a
diminishing difference between the inverse effective mass $m_0/m^*$ 
and the wave function renormalization $Z_{\vec k = 0}$
as $D$ increases.  The DMFT approach thus 
gives better results in higher dimensions.
Our comparison shows that DMFT
gives qualitatively correct results for the effective mass, mean
phonon number, and on-site phonon distortion in the intermediate- to
strong-coupling regime.  We also examine the comb-basis approach which limits
the el-ph correlation to the on-site level as DMFT does. The discrepancy
between the comb basis and the full basis decreases slowly as D increases.

Finally, our approach is compared to the well-known Toyozawa variational
method. We quantitatively examine the method in the intermediate-coupling
regime. Overall, the Toyozawa wavefunction gives reasonably accurate energy
and 2-point functions, but fails seriously for the 3-point functions. 
(The numerically exact 3-point functions are quite different
for excitations on opposite sides of the electron compared
to the same side, whereas the Toyozawa wavefunction predicts
that they are identical.)
We propose an improved variational wavefunction, a sum
of two asymmetric phonon clouds (Eq.\ \ref{eq:newTY}),
which gives improved 3-point functions, and
somewhat more accurate results for the energy, 
Z-factor, and 2-point el-ph correlation functions.

For all the polaron features calculated, the present numerical
approach compares favorably to other numerical methods in terms of
accuracy, ease of implementation, and the ability to compute ground
and excited state energies and correlation functions. 
It can also be directly applied to study
the effects of dimensionality on other interesting problems, 
such as the Fr\"ohlich
model or extended Holstein model with longer range
electron-phonon interactions, and to bipolaron problems.

The authors are grateful to S.~Ciuchi, E.~Jeckelmann, and P.~Kornilovitch
for discussions and permission to use their data,
and to J.~E.~Gubernatis and K.~K.~Loh for valuable discussions.
This work was supported by the U.S. Department
of Energy and by Los Alamos LDRD.


\begin{table}

\begin{tabular}{lllll} 
   &  1D & 2D   & 3D  & 4D \\ \tableline
 $E_0$  & -2.46968472393287071561  & -4.814735778337  &  -7.1623948409 &
 -9.513174069 \\
\end{tabular}
\vskip .5cm
\caption {Polaron ground state energies at
$k=0$ in 1D - 4D for $\alpha=0.5,~ g=1.0$, and $t=1.0$ .}

\vskip 2cm
\begin{tabular}{cllllll}  
       &  $ E_0$  & $ \alpha _ 2 (0) $ &
       $ \alpha _ 3 (1,2) $ &
       $ \alpha _ 3 (1,-2) $  & $Z_{k=0}$ &   \\
\tableline
 this work & -2.69356579774920\ldots & 0.40770 & 0.0004691 & 0.000005888 &  0.627322\ldots & \\
 Toyozawa &  -2.662819 &   0.32527  & 0.0002142 & 0.0002142 & 0.65738 & \\
 Eq.\ \ref{eq:newTY}& -2.671530 & 0.34240  & 0.0007649 & 0.000003244& 0.64271 &\\
 Shore-Sander & -2.685826 & 0.37780 & 0.0005572 & 0.0001132 & 0.63757& \\
 \end{tabular}
 \vskip .5cm
\caption[long]{A comparison of the ground state energy $E_0$, 
two- and three-point el-ph correlation
functions, and $Z_{k=0}$, evaluated by the present method, the Toyozawa,
Eq.~\ref{eq:newTY}, and Shore-Sander wavefunctions 
($\Psi^{IV}$ in Ref.~\cite{toyo3}). 
Parameters are $\lambda = 1.2$, $\omega = 1$, $t = 1$, $D=1$.}

\newpage
\vskip 2cm

\begin{tabular}{ccc}
   site $j$ & $z_j$ &\\ \tableline
  -6  & -0.12384D-03   & \\
  -5  & -0.39019D-03   &\\
  -4  & -0.11875D-02   & \\
  -3  & -0.32178D-02   & \\
  -2  & -0.48444D-02   & \\
  -1  &  0.35290D-01   &\\
   0  &  0.58515D+00   &\\
   1  &  0.38153D+00   &\\
   2  &  0.14043D+00   &\\
   3  &  0.46112D-01   &\\
   4  &  0.14632D-01   &\\
   5  &  0.45908D-02   &\\
   6  &  0.14349D-02   &\\
\end{tabular}
\vskip 1cm
\caption{ A partial list of the optimized phonon wavefunction 
$z_j$ in Eq.\ \ref{eq:newTY}.  
            }

\end{table}

\begin{figure}

\vskip .7in  \centerline{
\psfig{figure=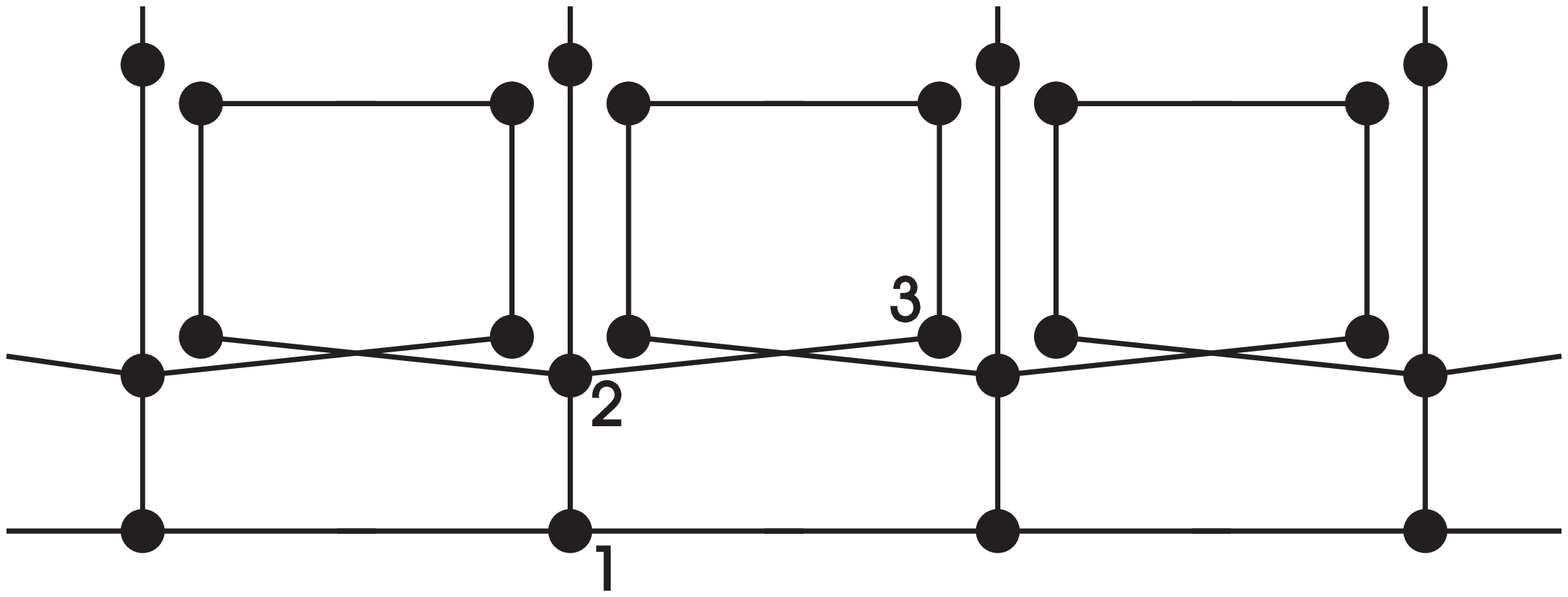,height=6cm,width=16cm,angle=0} } \vskip 1cm
\caption{A small variational Hilbert space, a subset of
the generation 3 space, is shown for the 1D polaron.
Basis states in the many-body Hilbert space are represented
by dots, and nonzero off-diagonal matrix elements by lines.
State $|1 \rangle$ in the root state, an electron at
the origin with no phonon excitations.  Vertical bonds
create or destroy phonons.  State $|2 \rangle$ is
an electron and one phonon, both at the origin.
State $|3 \rangle$ is an electron on site 1, and a phonon on site 0.
The dots can also be thought of as Wannier orbitals in a one-body
periodic tight-binding model.
\label{fig:tight}}

\newpage
\vskip .7in  \centerline{
\psfig{figure=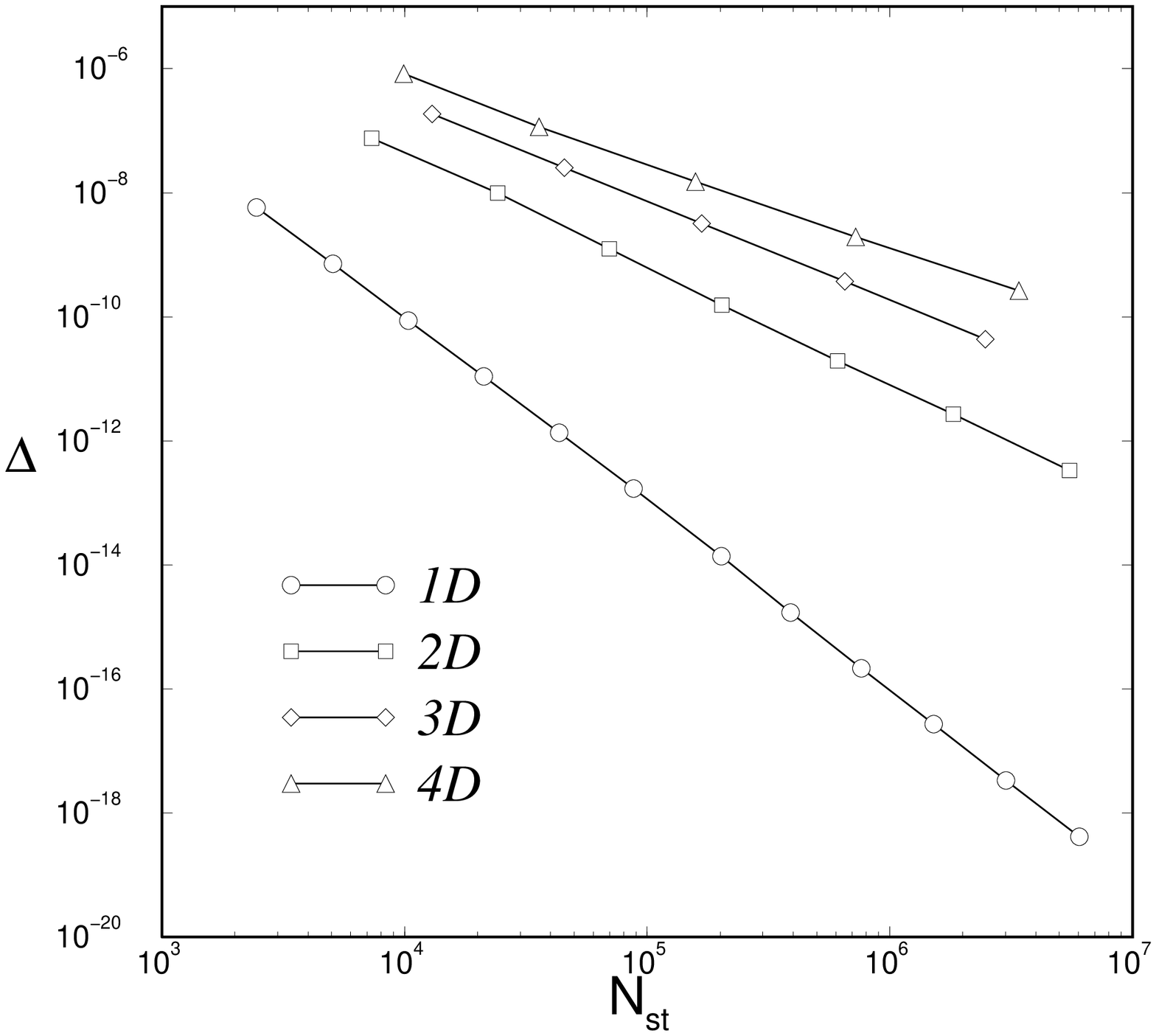,height=16cm,width=16cm,angle=-90} } \caption{The
fractional error $\Delta$ in the polaron ground state energy
as a function of the number of basis states $N_{st}$
in the Hilbert space for parameters 
$\alpha=0.5$, $g=1.0$, and $t=1.0$. \label{fig:convg}}

\newpage
\vskip .1in  \centerline{
\psfig{figure=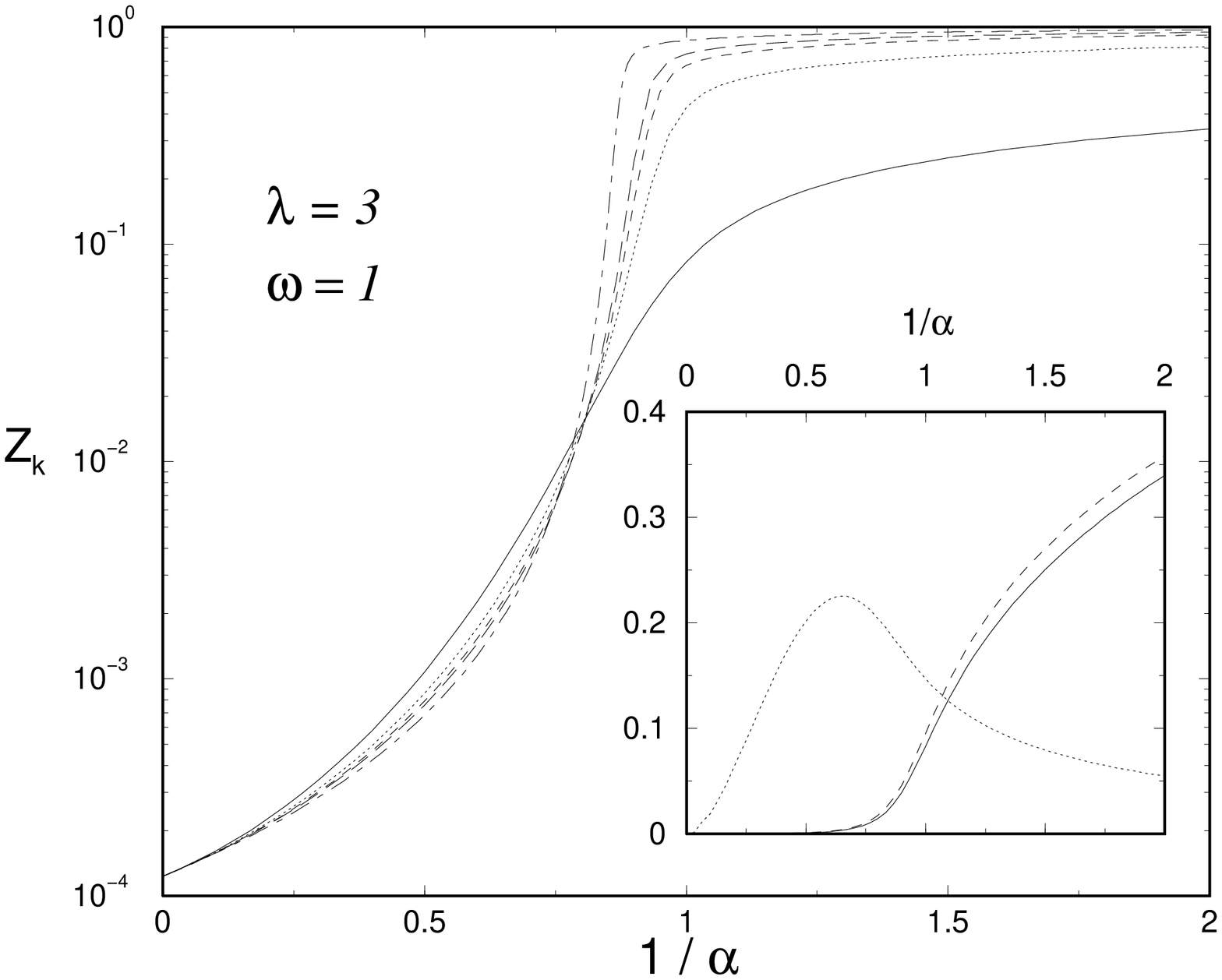,height=16cm,width=16cm,angle=-90} } \vskip 1cm
\caption{Quasiparticle weight $Z_{\vec k=0}$ as a function of the inverse
coupling strength $1 / \alpha$ for 1D (solid line), 2D (dotted line), 3D
(dashed line), and 4D (long dashed line).  $\alpha$ is varied 
by changing the hopping $t$ at fixed $\omega$ and $\lambda$.
The comb basis approximation (see below) is shown as a dot-dashed line.
The inset shows the comparison of
the inverse effective mass $m_0/m^*$ (dashed line) 
and $Z_k$ (solid line) for 1D. 
The fractional difference
$\delta= (m_0/m^* - Z_k)/Z_k$ is plotted
as a dotted line. \label{fig:crossover}}

\newpage
\vskip 2cm \centerline{ \psfig{figure=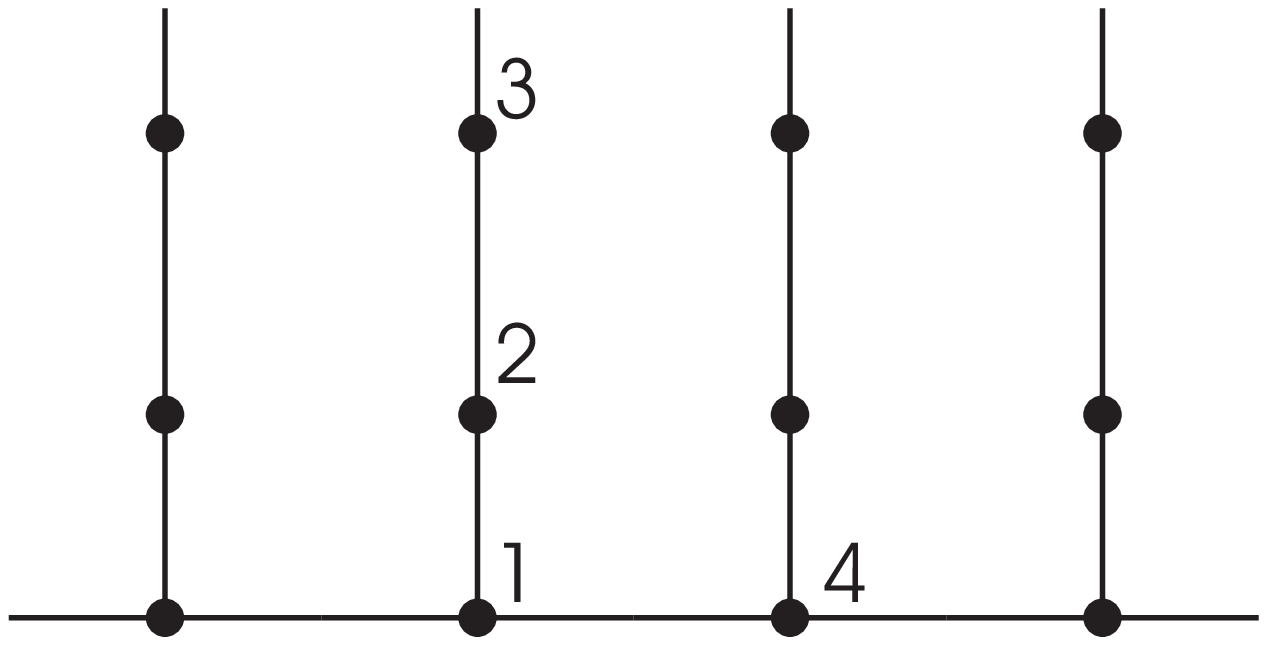,height=7.5cm,width=15cm,angle=0}
} \vskip 0.9in \caption{The comb basis, a variational space
in which phonon excitations are present only on the electron site.
Vertical lines create phonons and horizontal lines are the electron
hops.  State $|1\rangle$ is an electron on site 0 and no phonons. State
$|2\rangle$ is an electron and one phonon, both on site 0. State $|3\rangle$ is
an electron and two phonons, all on site 0. State $|4\rangle$ is the a
translation of state $|1\rangle$.  The comb basis is a subset of the 
larger variational
space.  As in DMFT, it only keeps track of the on-site el-ph correlations.
\label{fig:comb}}

\newpage
\vskip .7in  \centerline{
\psfig{figure=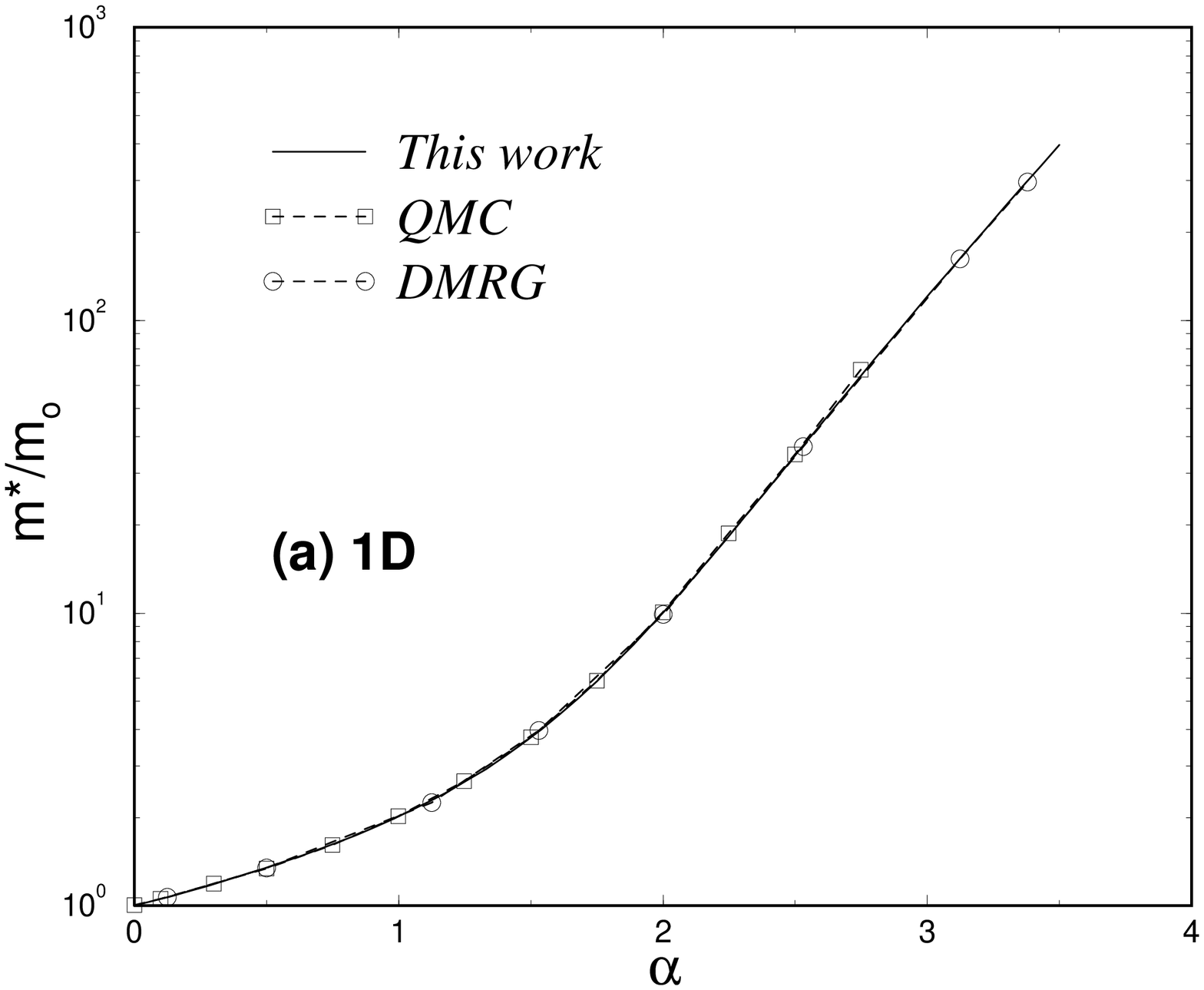,height=16cm,width=16cm,angle=-90} } \vskip 1cm
\caption[long]{ The effective mass $m^*$ for the (a) 1D, (b) 2D, and (c) 3D
polaron is compared to DMRG \cite{DMRG} and QMC \cite{korn98} calculations.
(No DMRG
data is available for 3D polaron.)  In all cases, $\omega=1.0$, and 
$t = 1.0$.  Note different horizontal scales. \label{fig:DMRG}}

\newpage
\vskip .7in  \centerline{
\psfig{figure=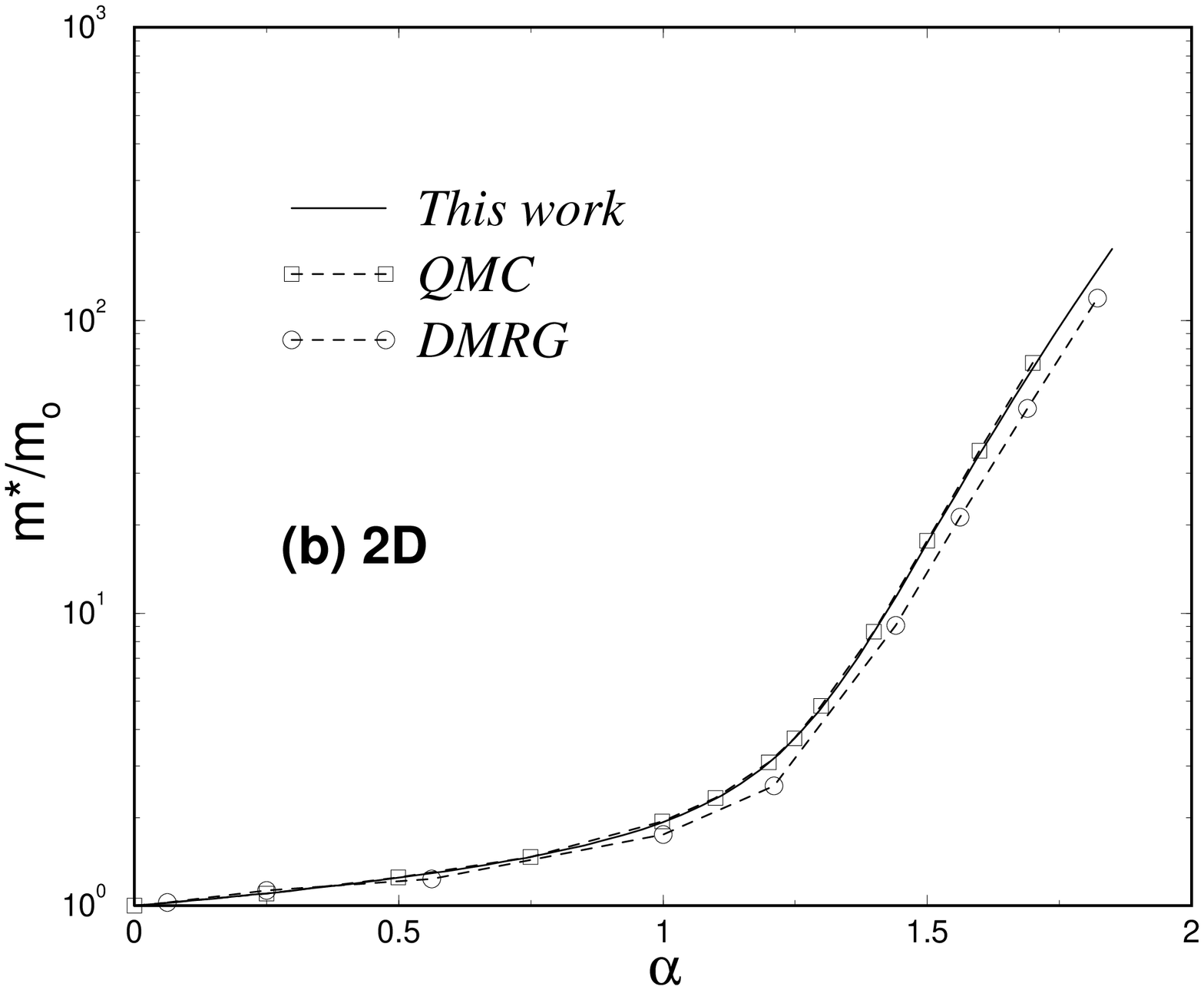,height=16cm,width=16cm,angle=-90} } \vskip 1cm

\newpage
\vskip .7in  \centerline{
\psfig{figure=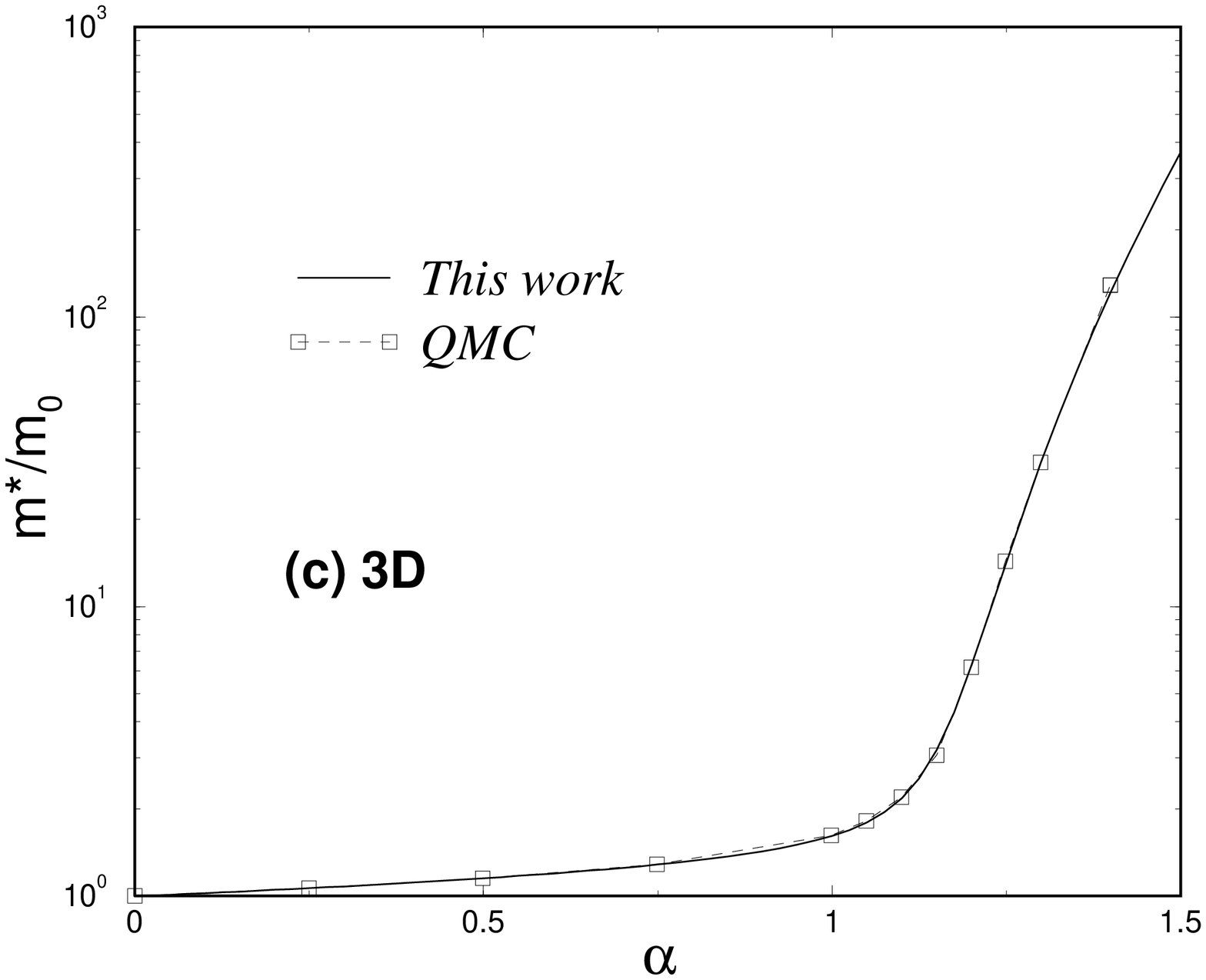,height=16cm,width=16cm,angle=-90} } \vskip 1cm

\newpage
\vskip .7in  \centerline{
\psfig{figure=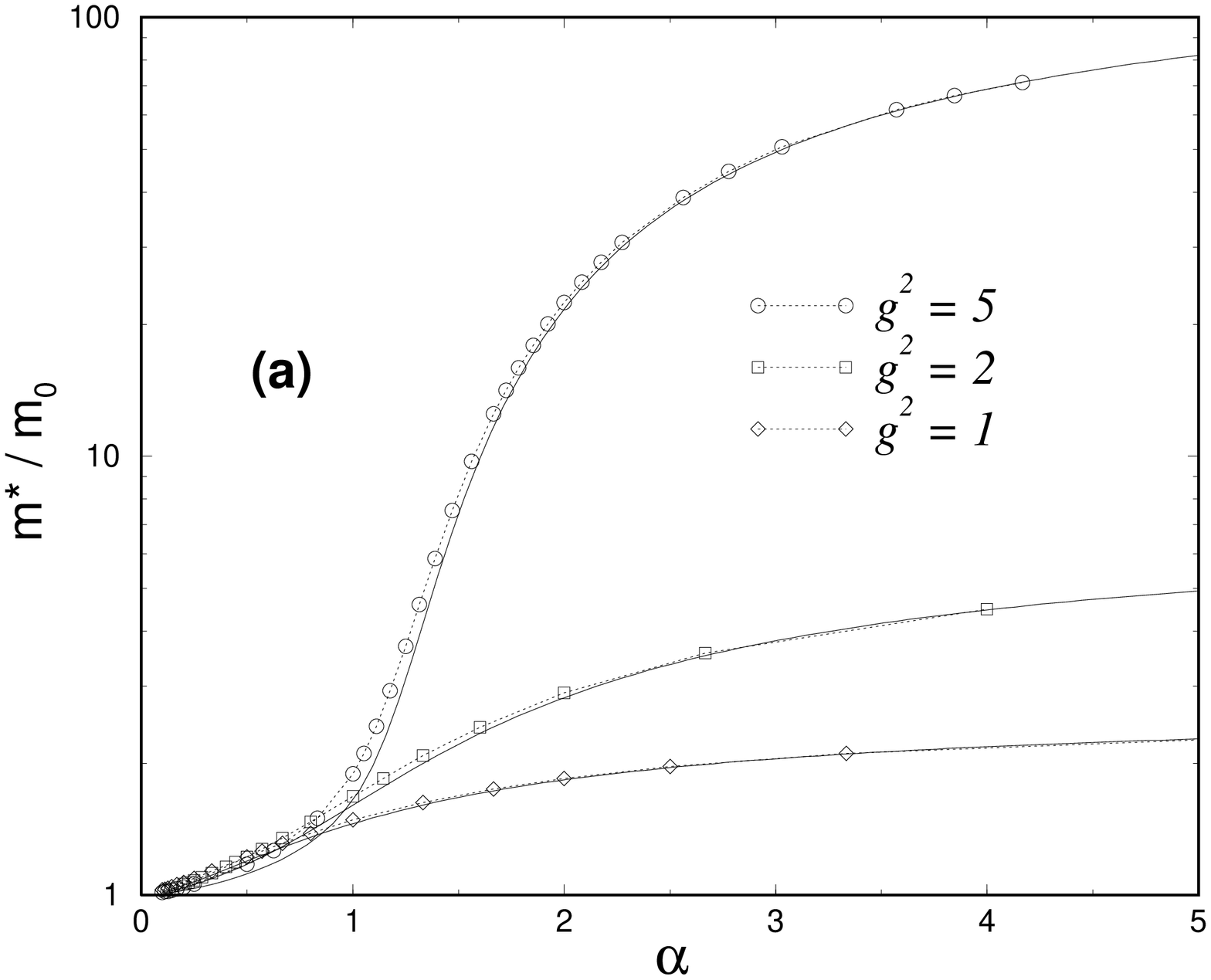,height=16cm,width=16cm,angle=-90} } \vskip 1cm
\caption[long]{(a) : The mass $m^*$ of the 3D polaron,
$\omega = 1$, this work (solid lines)
compared to DMFT (dotted lines)  \cite{DMFT}. 
(b) : Comparison to weak-coupling perturbation theory (WCPT) for
$g^2 = 5$.
\label{fig:DMFT} }

\newpage
\vskip .7in  \centerline{
\psfig{figure=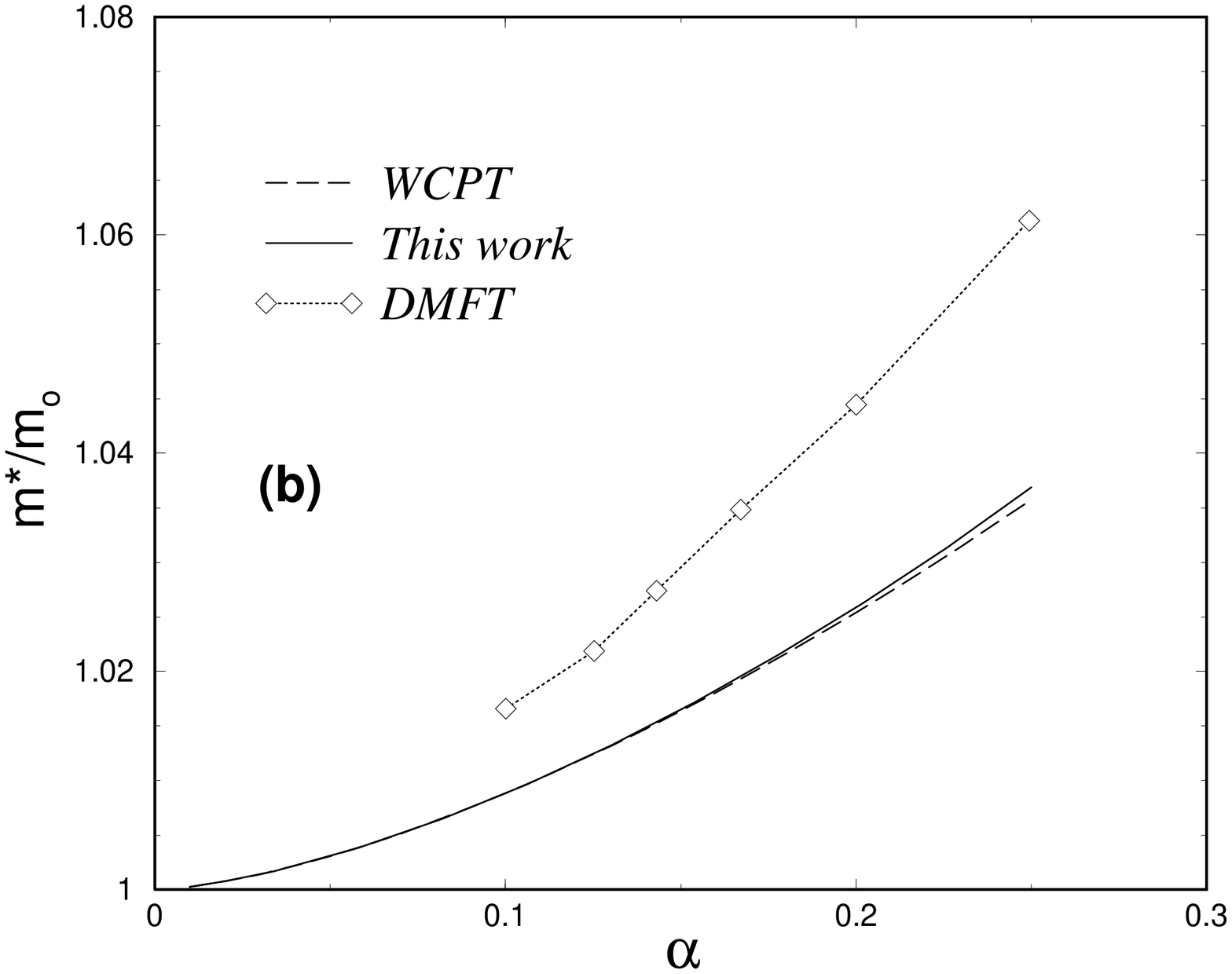,height=16cm,width=16cm,angle=-90} } \vskip 1cm

\newpage
\vskip .7in  \centerline{
\psfig{figure=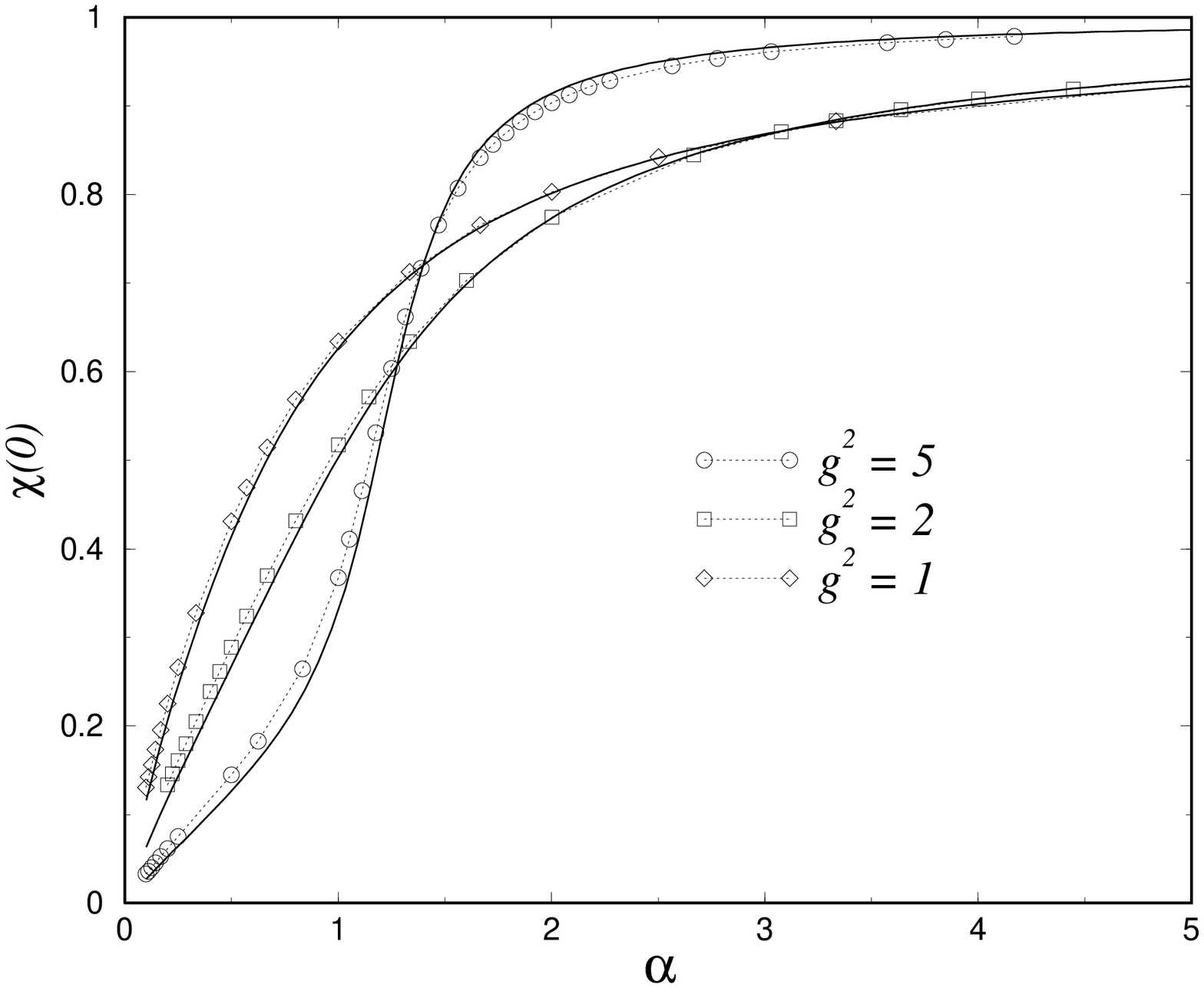,height=16cm,width=16cm,angle=-90} } \vskip 1cm 
\caption{ The on-site correlation $\chi(0)$ for the 3D polaron. Our results
(solid lines) are compared to DMFT (dotted lines with symbols). The parameters 
are the same as in Fig.~\ref{fig:DMFT} \label{chi0} }

\newpage
\vskip .7in  \centerline{
\psfig{figure=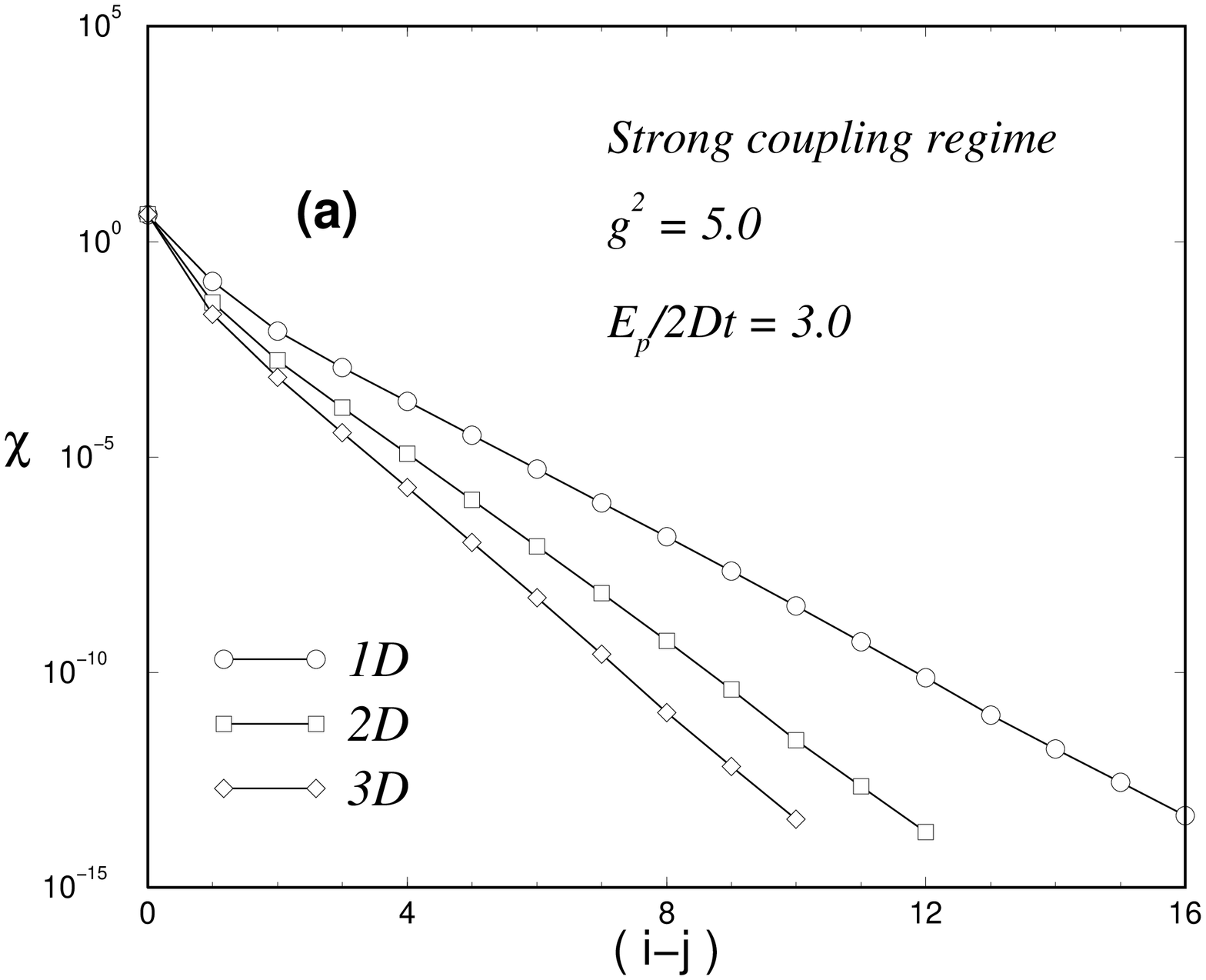,height=16cm,width=18cm,angle=-90} } \vskip 1cm
\caption{ Correlation $\chi$ of the electron density and the phonon
displacement
as a function of distance $(i-j)$ for the 3D polaron
along the ($1,0,0$) direction, the 2D polaron along the ($1,0$)
direction, and the 1D polaron at (a) strong coupling, (b) weak
coupling, $\omega=1.0$. 
Note the different vertical scales.  \label{fig:deform} }

\newpage
\vskip .7in  \centerline{
\psfig{figure=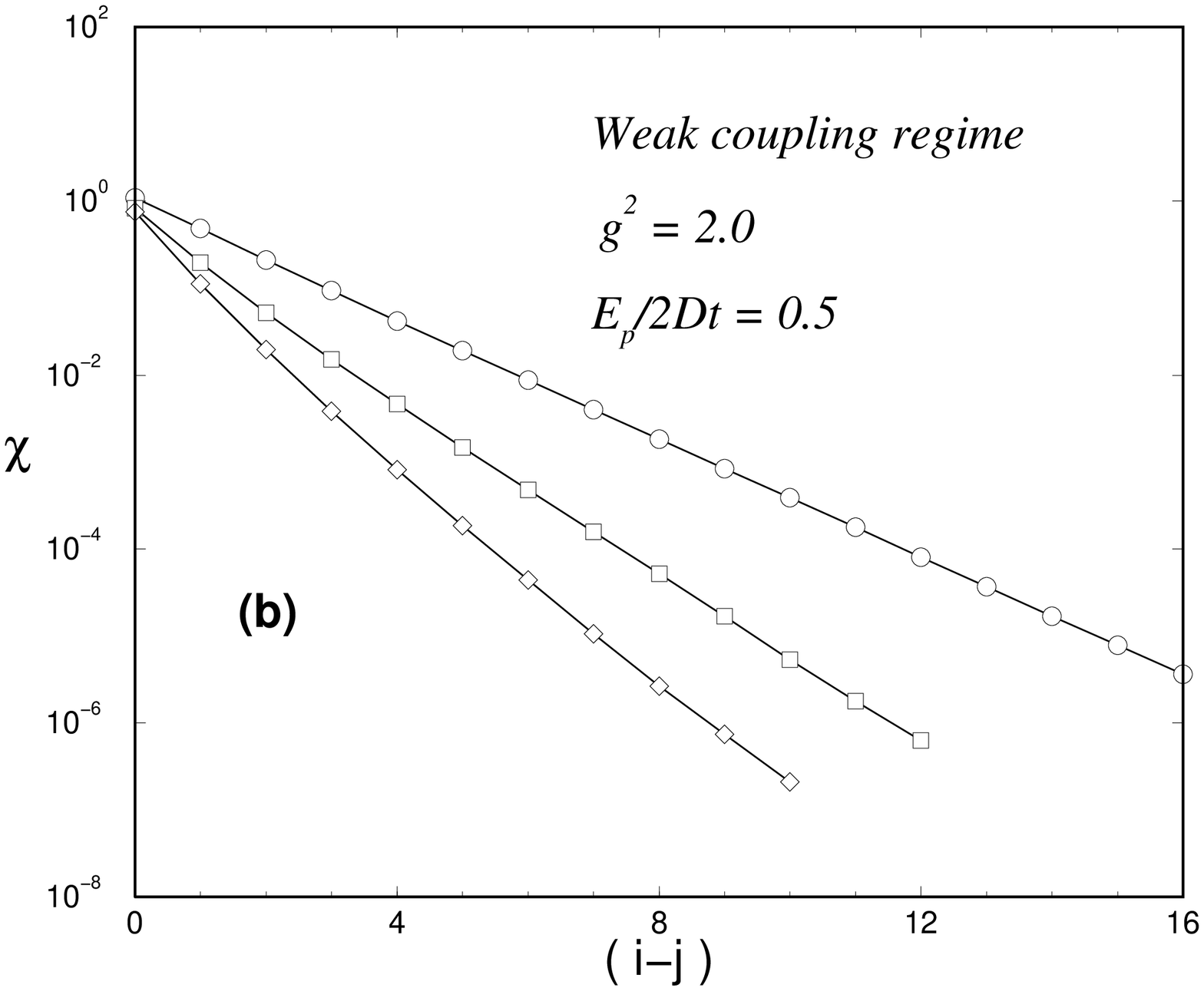,height=16cm,width=18cm,angle=-90} } \vskip 1cm

\newpage
\vskip .7in  \centerline{
\psfig{figure=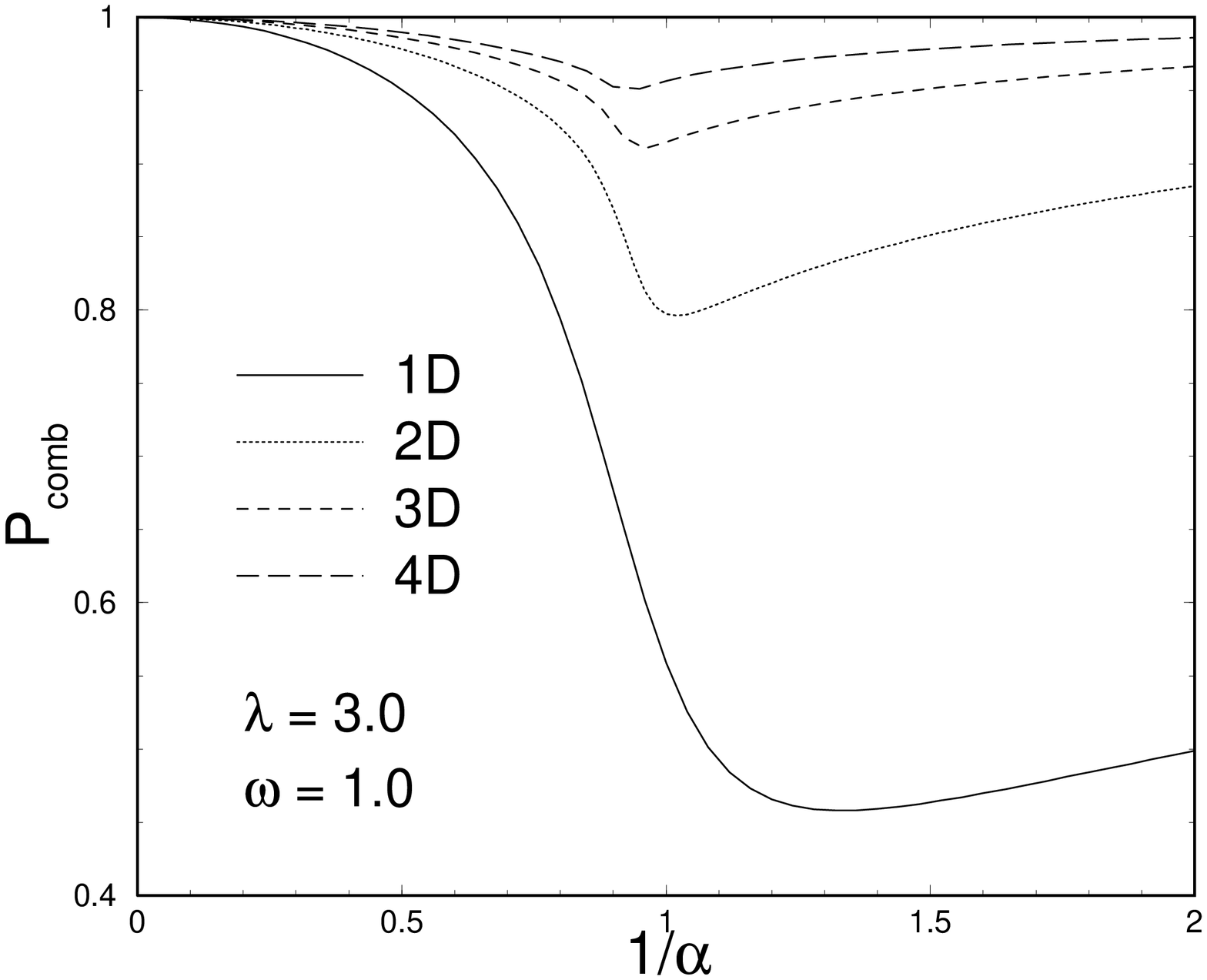,height=16cm,width=18cm,angle=-90} } \vskip 1cm
\caption{ The probability density in the ground state that resides in
the comb subspace $P_{comb}$ as a function of the
inverse bare coupling strength $1/\alpha$, for the 1-4D polaron. The parameter
set is the same as in Fig.\ \ref{fig:crossover}. \label{fig:P_comb}}

\newpage
\vskip .7in  \centerline{
\psfig{figure=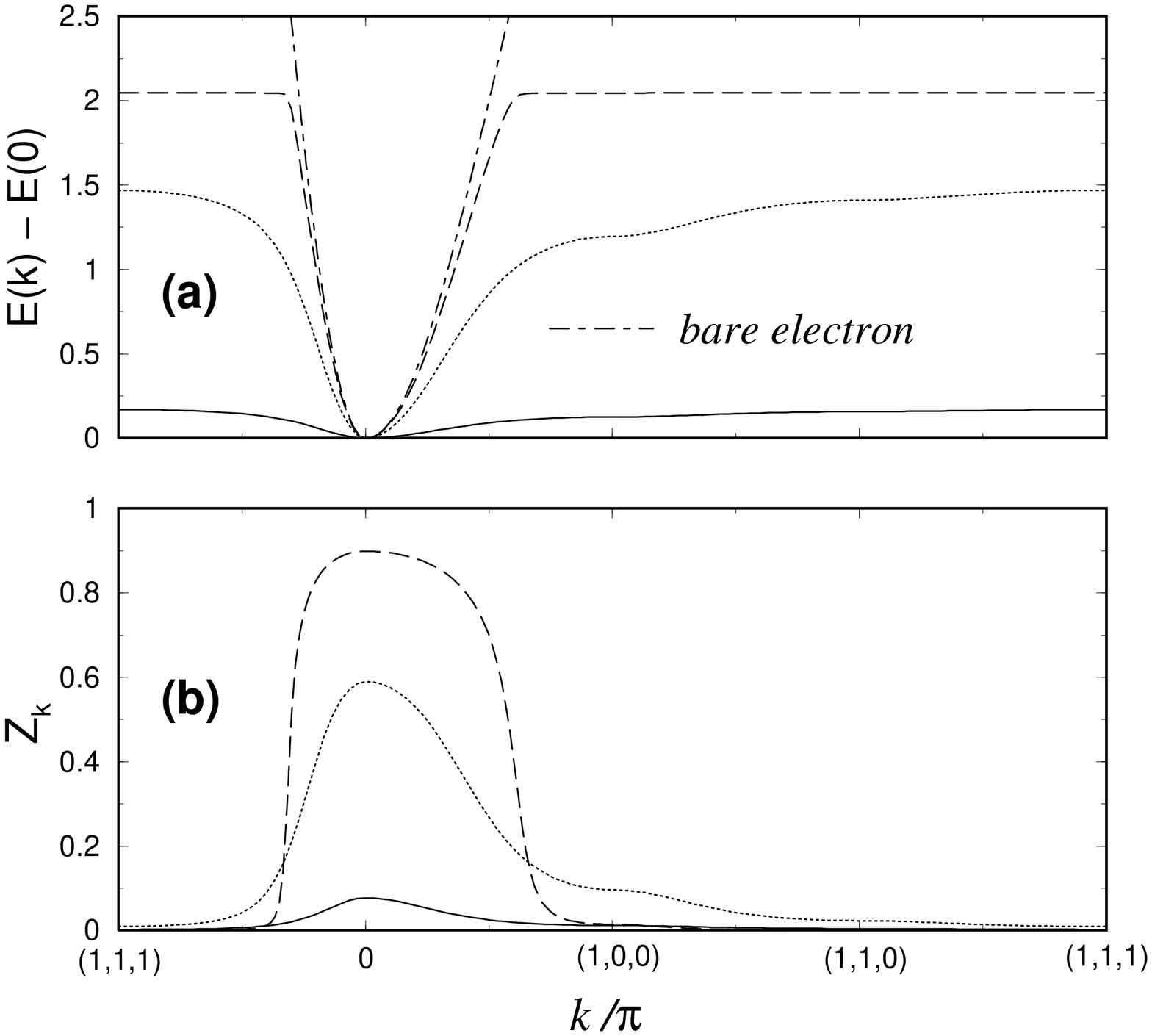,height=16cm,width=17cm,angle=-90} } \vskip 1cm
\caption{Ground state energy $E(\vec k )$ of the 3D polaron in panel (a) and
quasiparticle weight $Z_k$ in panel (b) for three different el-ph coupling
constants, $\lambda=4.5$ (solid line), $\lambda=3.5$ (dotted line),
and $\lambda=2.0$ (dashed line). Other parameters are $\omega = 2.0$ and $t =
1$.  The dot-dashed line in (a) is the dispersion of a bare electron. 
The corresponding ground state energies $E(\vec k =0 )$ are 
$-10.608348,\; -8.0642850$, and $-6.588526818$ respectively. \label{fig:band} }

\newpage
\vskip .7in  \centerline{
\psfig{figure=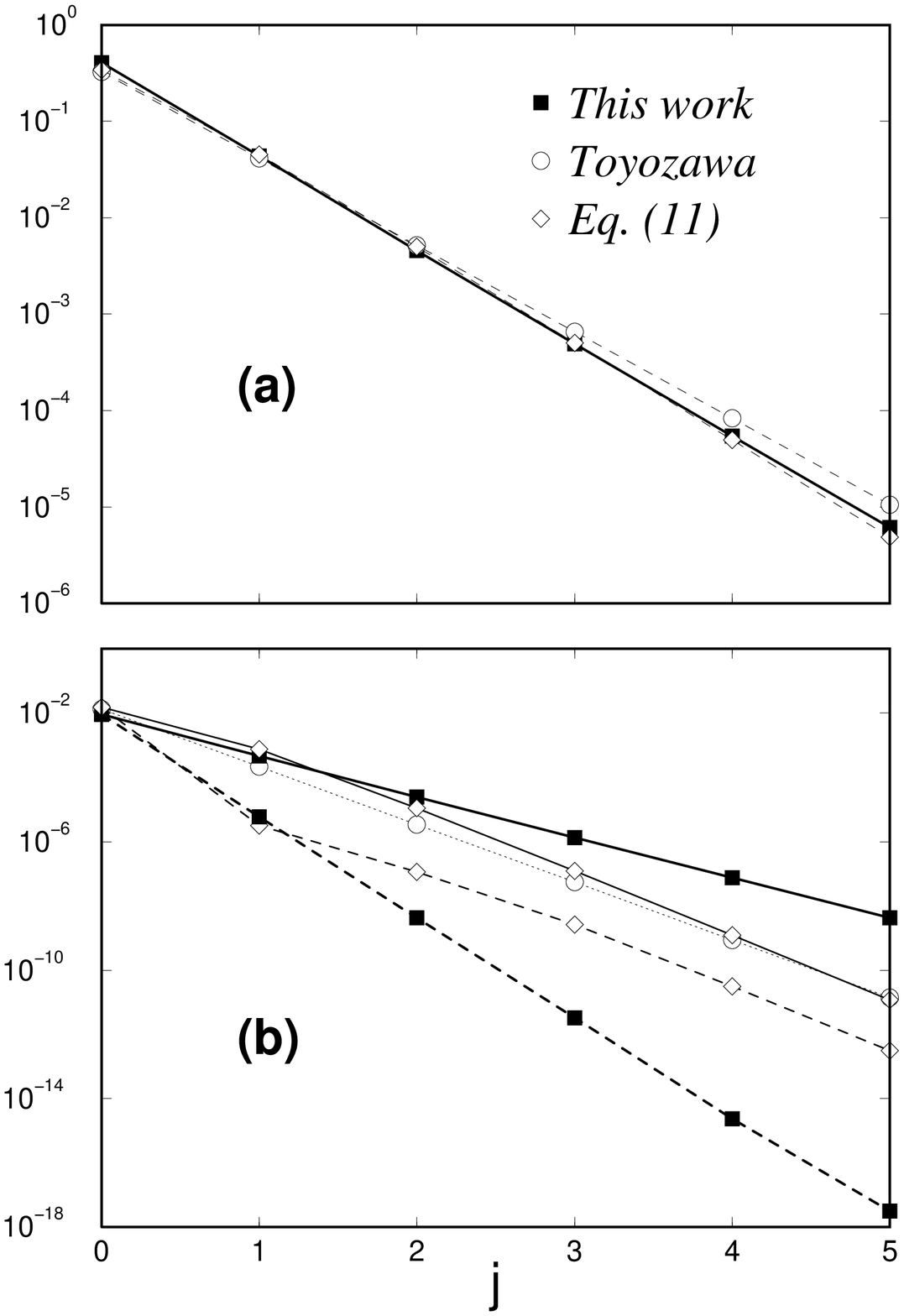,height=15cm,width=11cm,angle=0} }    \vskip 1cm
\caption{ (a) The 2-point function $\alpha_2 (j)$ is evaluated 
in 1D by the present variational method (solid line with squares),
the Toyozawa method (dashed line with circles), and the modified
Toyozawa method Eq.\ \ref{eq:newTY} (dashed line with diamonds). (b) The
3-point functions $\alpha_3(j,j+1)$ (solid lines) and $\alpha_3(j,-j-1)$
(dashed lines). The symbols are the same as in (a). Note that
the plain Toyozawa method gives exactly the same results for the two 3-point 
functions, which in fact differ widely. 
Parameters are $\omega = 1.0,~ t=1.0,$ and $\lambda =
1.2$. \label{fig:Tyz} }

\newpage
\vskip .7in  \centerline{
\psfig{figure=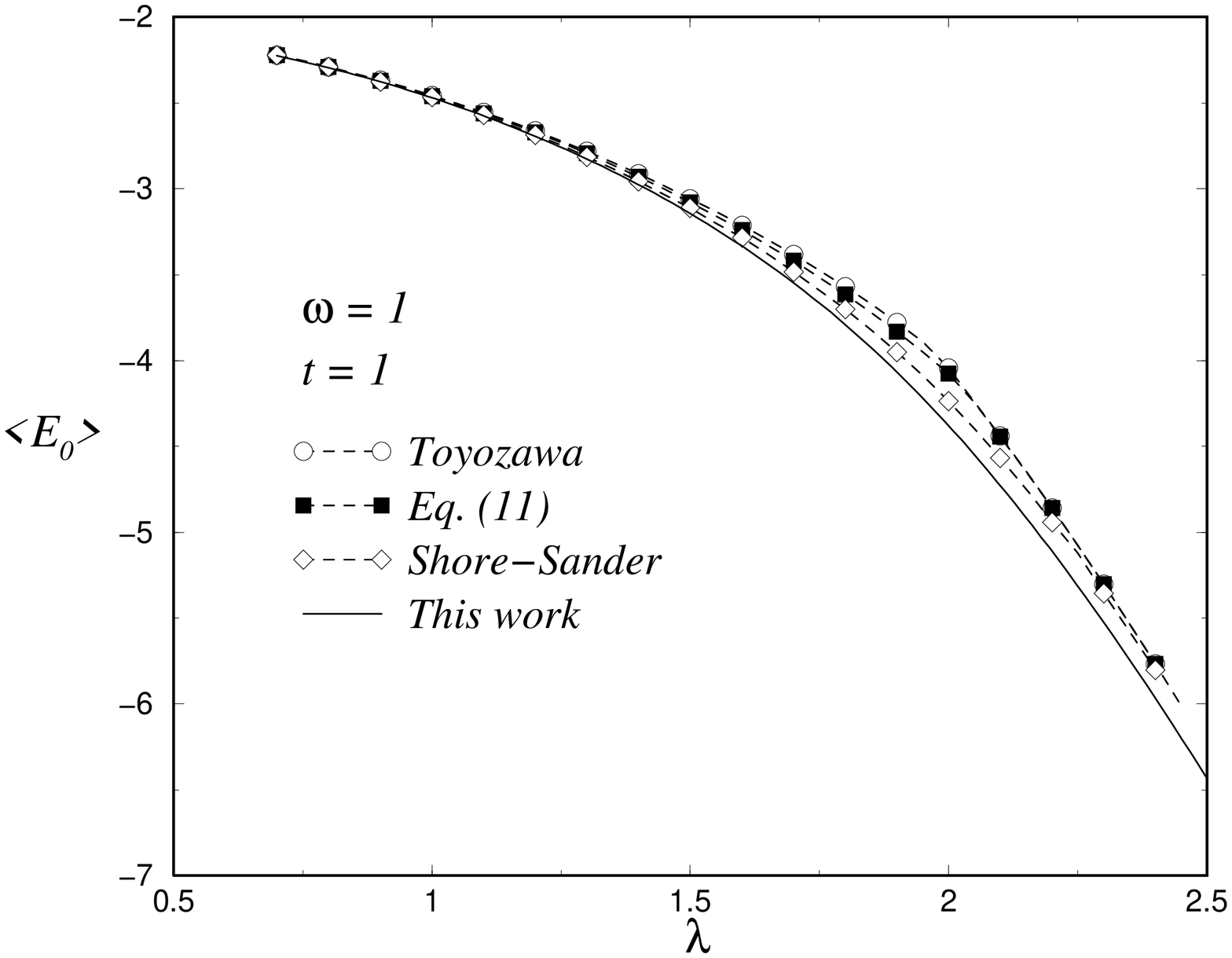,height=16cm,width=14cm,angle=-90} } \vskip 1cm
\caption{A comparison of ground-state energy as a function of coupling 
constant from various variational approaches for $\omega=1, t=1$.
\label{fig:TY_Shore}}


\end{figure}


\begin{references}
\bibitem{com1} Problems with other types of electron-phonon coupling and lattices,
as well as the two electron (bipolaron) problem can be solved by similar methods.
\bibitem{strongc}  For strong electron-phonon coupling, it is sometimes
advantageous to add more phonon basis states very near to the electron,
called a ``tower''.  Even with a tower, convergence is slower
at strong coupling, corresponding to a shallower slope in 
Fig. \ref{fig:convg}.
\bibitem{norm}  All presented el-ph correlation functions are
evaluated for wavefunctions that are normalized to one electron per site.
For conventional normalization, there would be an additional $1/N$ factor,
where $N$ is the number of sites in the system.
\bibitem{accuracy} The error in the flattened band is about $2.1\%$ for the 3D
polaron. In the weak-coupling regime, the extent of the 
el-ph correlations at $k=\pi$
is much larger than at $k=0$.
For the 1D polaron, the el-ph correlations decay exponentially
at all $k$ for all parameters; the polaron and additional phonon
are bound at $k = \pi$.  It is not clear that this
is the case in 3D.
\bibitem{linden}  A comparison of energies
obtained by a wider variety of methods is contained in 
Ref.~\cite{GL}.  Note that what is referred to as ``Toyozawa variation''
in that reference is a more sophisticated trial
function than what we call the Toyozawa wavefunction,
Eq.\ \ref{eq:TY}.
\end{references}
\end{document}